\documentstyle[12pt]  {ioplppt}
\begin{document}
\jl{1}
\letter{Logarithmic corrections to gap scaling in random-bond Ising strips}
\author{S L A de Queiroz\ftnote{1}{E-mail: sldq@if.uff.br}}

\address{Instituto de F\'\ii sica, UFF, Avenida Litor\^anea s/n,
Campus da Praia Vermelha, 24210--340 Niter\'oi RJ, Brazil}

\begin{abstract}
Numerical results for the first gap of the Lyapunov spectrum of the self-dual random-bond Ising model
on strips are analysed. It is shown that finite-width corrections can be
fitted very well by
an inverse logarithmic form, predicted to hold when the Hamiltonian contains
a marginal operator. 
\end{abstract}

\maketitle

\nosections
It is widely believed that the critical behaviour of the two-dimensional
Ising model is only slightly modified by the introduction of non-frustrated
disorder~\cite{sst,sbl}. Such changes are given by logarithmic corrections
to pure-system power-law singularities. In terms of a field-theoretical (or renormalisation-group) description, disorder is said to be a marginally 
irrelevant operator~\cite{ludwig,shalaev}. 
Specific forms have been proposed, and 
numerically tested, for the corrections to bulk quantities  as specific
heat, magnetisation and initial susceptibility~\cite{sst}. The overall
picture emerging from such analyses tends to confirm predictions of
log-corrected pure-Ising behaviour. Early proposals of
drastic alterations in the values of critical indices~\cite{dd} have thus been
essentially discarded. However, recent results~\cite{kim,kuhn} have appeared,
according to which critical indices would vary
with disorder, but so as to keep the ratio $\gamma/\nu$ constant at the
pure system's value  (the so-called {\it weak universality} 
scenario~\cite{suzuki}). In order to solve this controversy, it is important
to undertake independent tests of several aspects of the problem. In the
present work we examine the finite-width corrections to the first gap of
the Lyapunov spectrum of the disordered Ising model on strips. 
This gap is related to the typical (as opposed to averaged) behaviour
of spin-spin correlation functions, as explained below.

While the average of a random quantity ${\cal Q}$ is simply its
arithmetic average over independent realisations ${\cal Q}_i$, 
$\overline{{\cal Q}} = (1/N)\sum_{i=1}^N {\cal Q}_i$, its typical (in the
sense of most probable) value is expected to agree with
the geometrical average:
${\cal Q}_{typ} = {\cal Q}_{m.p.} = \exp\left(\overline{\ln {\cal
Q}}\right)$~\cite{dh,derrida,crisanti,ranmat}. 
Depending on the underlying probability distribution, 
$\overline{{\cal Q}}$ and ${\cal Q}_{typ}$ may differ appreciably, as is
the case when one considers correlation functions~\cite{dh,crisanti}.

It has been predicted, on the basis of field-theoretical arguments~\cite{ludwig},
that in a bulk system (as opposed to the strip geometry used here) the {\it
typical} decay of spin-spin correlation functions at criticality on
a fixed sample is given by  
\begin{equation}
 \langle \sigma_0 \sigma_R \rangle_{typ} \propto R^{-1/4} (\Delta \ln R)^{-1/8}
\ \
 \  {\rm (fixed\ sample)}
\label{eq:1}
\end{equation}
for $\ln (\Delta \ln R)$ large, where $\Delta$ is proportional to the intensity
of disorder. This way, pure-system behaviour (power-law decay against distance,
$\langle \sigma_0 \sigma_R \rangle \sim R^{-\eta}$ with $\eta=1/4$) 
acquires
a logarithmic correction. On the other hand, if one considers the {\it average},
over many samples, of the
correlation function, logarithmic corrections are expected to be washed 
away~\cite{ludwig} resulting in a simple power-law dependence:  
\begin{equation}
  \overline{\langle \sigma_{0} \sigma_{R} \rangle} \propto R^{-1/4} \ \ \ 
{\rm (average\ over\ samples) .}
\label{eq:2}
\end{equation}
The distinction between typical and average correlation decay
was not explicitly discussed in the early field-theoretical treatment~\cite{dd}
which predicted $\langle \sigma_0 \sigma_R \rangle \propto e^{-A(\ln R)^2}$,
a sort of behaviour for which no evidence
has been found in subsequent investigations~\cite{sst}.
Recent numerical work claiming weak universality to hold~\cite{kim,kuhn}
does not address the issue either, though it is easy to see that 
the procedures used in those calculations pick out averaged correlations,
as they rely respectively on variants of the fluctuation-dissipation 
theorem~\cite{kim} or on explicit averaging~\cite{kuhn}.

In Ref.~\cite{kuhn}, an analysis of phenomenological renormalisation estimates of the correlation-length exponent $\nu$ seems to
point against logarithmic corrections to its pure-system value, and in
favour of a disorder-dependent exponent. However, that is based on data
for very narrow strips ($L \leq 8$) and mostly relies on trends apparently 
followed at weak disorder. Though one series of moderately-strong 
disorder results is exhibited as well, no attempt is made to fit
either set to the form predicted by theory, which crucially includes a
disorder-dependent crossover length~\cite{sst}.  
Indeed, a systematic treatment of the
averaged correlation lengths at the exact critical point, spanning a wide
range of disorder and watching for the disorder-dependent crossover mentioned
above, eventually uncovers the expected logarithmic terms~\cite{sbl}.                                                                                                                                                                                                                                                                        
 Turning back to the
exponent $\eta$, recall that, as far as dominant behaviour
is concerned, both weak- and strong-universality concur in predicting 
$\eta = 1/4$. Thus, our strategy here will be to analyse the {\it subdominant}
terms. Our goal is to show that  the finite-size corrections to the typical
ratio of decay of correlations behave consistently with what is expected
when a marginal operator is present (see below).    

In numerical simulations, one considers finite lattices
($L \times L$) or finite-width strips ($L \times N$, $N \to \infty$) and sets
the temperature at the critical point of the corresponding two-dimensional
system. For the nearest-neighbour random-bond Ising model on a square lattice,
with a binary distribution of ferromagnetic interactions
\begin{equation}
 P(J_{ij})= {1 \over 2} ( \delta (J_{ij} -J_0) +  \delta (J_{ij} -rJ_0) ) \ \ \ ,\ \ 0 \leq r \leq 1 \ \ ,
\label{eq:3}
\end{equation}
the critical temperature $\beta_c = 1/k_B T_c$ 
is exactly known~\cite{fisch,kinzel} from self-duality as a function of $r$ through:
\begin{equation}
\sinh (2\beta_{c} J_{0})\sinh (2\beta_{c}r J_{0}) = 1 \ \ .
\label{eq:4}
\end{equation}
From Monte Carlo work on $L \times L$ random-bond systems~\cite{talapov}, 
it has been found that the average correlation function
at criticality is numerically very close to the exactly-known~\cite{wu}
value for a pure system of the
same size at its own critical point, thus providing evidence in favour
of \Eref{eq:2}. Similar conclusions have been drawn for the corresponding
quantities evaluated on strips~\cite{dq,dqrbs}, where the exact critical
correlation functions for the pure Ising model are known from conformal
invariance~\cite{cardy}.

Here we provide a test of the consequences of \Eref{eq:1}, 
when correlations are considered on a strip geometry.
In this case, contrary to that of \Eref{eq:2}, no exact finite--$L$
results are available for comparison; thus one must resort to finite-size scaling concepts~\cite{fs1}, in order to unravel signs of the expected bulk behaviour from trends followed by finite-system data as $L \to \infty$.

The procedure used is as follows. It is known that the typical, or most
probable, spin-spin correlation function on a strip decays as 
\begin{equation}
 \langle \sigma_0 \sigma_R \rangle_{typ} \propto \exp\left(-R/\xi_{typ}\right),
\ \ \xi_{typ}^{-1} = \Lambda_L^0 - \Lambda_L^1\ ,
\label{eq:5}
\end{equation}
where $\Lambda_L^0$ and $\Lambda_L^1$ are the two largest Lyapunov 
exponents of the (random) transfer matrix for a strip of width
$L$~\cite{derrida,crisanti,ranmat}.
On the other hand, conformal invariance predicts~\cite{car86} that, when the
Hamiltonian of a homogeneous two-dimensional system contains a
marginal operator, the spectrum of eigenvalues $E_n$ of the transfer matrix
on a strip is such that
\begin{equation}
E_n - E_0 = (2\pi/L)(x_n + d_n/\ln L) + \ldots \ \ \ ,
\label{eq:6}
\end{equation}
where $x_n$ is the corresponding scaling dimension,
$d_n$ is an $n$-dependent constant and  periodic boundary conditions are
used across the strip.
Since $(i)$ disorder in the two-dimensional Ising model is believed to 
be a marginal operator, $(ii)$ Lyapunov exponents of transfer matrices in
random systems are the counterparts of eigenvalues in homogeneous ones, and
$(iii)$ numerical evidence shows that conformal-invariance results
derived for uniform systems may be extended to random cases, provided that
suitable averages are taken~\cite{sbl,dq}, we shall examine sequences of
estimates $\Lambda_L^0 - \Lambda_L^1$ for varying $L$ and try to fit them
to \Eref{eq:6} with $n=1$ and $2 x_1 = \eta = 1/4$~\cite{cardy}.    

We have used strips of width $L \leq 13$ sites, and length $N = 10^5$
for $L = 2 - 11$ and $5 \times 10^4$ for $L=12$ and $13$. Two values
were taken for the disorder parameter $r$ of \Eref{eq:3}: $r=0.5$ and $0.01$.     Details of the calculation are given in Ref. ~\cite{dq}, where the data
used here are displayed as well, in plots of
$(L/\pi)(\Lambda_L^0 - \Lambda_L^1)$ against $1/L^2$. The
latter variable was used because it is exactly known~\cite{dds} that, 
for strips of pure Ising spins,
the leading corrections to the $L^{-1}$
dependence of $\xi^{-1}$ are proportional to $L^{-2}$. 
Our purpose there was to show that
the correlation length $\xi_{ave}$ coming from direct average
of correlation functions, 
$\overline{\langle \sigma_{0} \sigma_{R} \rangle} \sim \exp(-R/\xi_{ave})$,
indeed scales as its pure-system counterpart, while $\xi_{typ}$ does not. At
the time we did not investigate the behaviour of $\xi_{typ}$ in detail,
though it was noticed that for strong disorder the curvature of plots of
$L/\pi\xi_{typ}$ against $L^{-\phi}$ only became smaller for $\phi$ close to
zero.

In Figures 1(a) and 1(b) we show 
$L/\pi\xi_{typ} \equiv (L/\pi)(\Lambda_L^0 - \Lambda_L^1)$ 
against $1/\ln L$, respectively for $r=0.5$ and $0.01$. For weak disorder
$r=0.5$ the fit to \Eref{eq:6} is very good, with $2 x_1 = \eta =1/4$.
While for $r=0.01$  strong fluctuations increase the amount of scatter, the
overall picture still is consistent with \Eref{eq:6}. Table 1 shows
results from least-squares fits of data in the range $L = 6 - 13$. This interval
was chosen in order to minimise the accumulated standard deviation $\chi^2$
per degree of freedom~\cite{dq}.
\begin{table}
\caption{
Results from least-squares fits to \protect{\Eref{eq:6}}, for $L=6-13$.}
\begin{indented}
\item[]\begin{tabular}{@{}llll}
\br
\crule{4}\\
$r$ & $\eta$& $d_1$ & $\chi^2$\\
\mr
0.5 & 0.250 $\pm$ 0.006 &  0.018 $\pm$ 0.013 &.021\\
0.01 & 0.29 $\pm$  0.04 &  0.38 $\pm$ 0.09 & .17 \\
Expected & 1/4 & --- &  ---\\
\br
\end{tabular}
\end{indented}
\end{table}
The extrapolated $\eta$ is expected to be universal as long as $r \neq 0$;
our error bar for $r=0.01$ indeed includes $\eta =1/4$,
though admittedly at the edge. On the other hand, $d_1$ clearly changes, 
increasing with disorder. The situation differs from that of the $q$-state
Potts model with $q=4$~\cite{car86,bn,nb} where there is no continuously 
tunable parameter, as marginality depends on $q-4$ being strictly zero.

As an additional check on the ideas developed above, we examined data for the
probability distribution of critical spin-spin correlation functions~\cite{dqrbs}
for $r=0.25$. For spin-spin distances $R = 5$ and 20, and strip widths
$L = 3, \ldots 13$ we picked the averages $\overline{\ln G(R)} \equiv 
\overline{\ln\left(\langle \sigma_0 \sigma_R \rangle\right)}$.
The quantity $\exp\overline{\ln G(R)}$ is expected~\cite{crisanti,ranmat}
to scale as $\langle \sigma_0 \sigma_R \rangle_{typ}$.
For each $L$ the slope of a two-point semi-log plot of  
$\exp\overline{\ln G(R)}$ gave an approximate value for
$1/\xi_{typ}$. From a plot of $L/\pi\xi_{typ}$ against $1/\ln L$
one gets $\eta \simeq 0.26$, $d_1 \simeq 0.03$, consistent with the above
values derived directly from Lyapunov exponents, and with the assumption that
$d_1$ varies continuously (and monotonically) against disorder.     

We have analysed numerical estimates for the first gap of the Lyapunov
spectrum of the self-dual random-bond Ising model
on strips. We have shown that finite-width corrections can be
fitted very well by
an inverse logarithmic form, predicted to hold when the Hamiltonian contains
a marginal operator. The present results contribute, albeit indirectly,
to the growing body of evidence favouring strong universality (that is,
pure-system behaviour with logarithmic corrections) in the two-dimensional
random Ising model~\cite{sst,sbl,ludwig,shalaev}. This is to be contrasted to
recent work~\cite{kim,kuhn}, according to which critical indices would vary
with disorder, but so as to keep the ratio $\gamma/\nu$ constant at the
pure system's value  (the so-called {\it weak universality} 
scenario~\cite{suzuki}).  

\ack
The author thanks CNPq, FINEP and CAPES for financial support.

\Figures

\begin{figure}
\caption[]{$\eta \equiv L/\pi \xi_{typ}$, with  $\xi_{typ}^{-1} = 
 \Lambda_L^0 - \Lambda_L^1$ of \protect{\Eref{eq:6}},
 against $1/\ln L$. The square on the vertical axis is at the pure system value $\eta=1/4$ . Straight lines are least-square fits for $L= 6-13$. $(a)$ : $r=0.5$; $(b)$ : $r=0.01$ .}
\label{fig:1}
\end{figure}

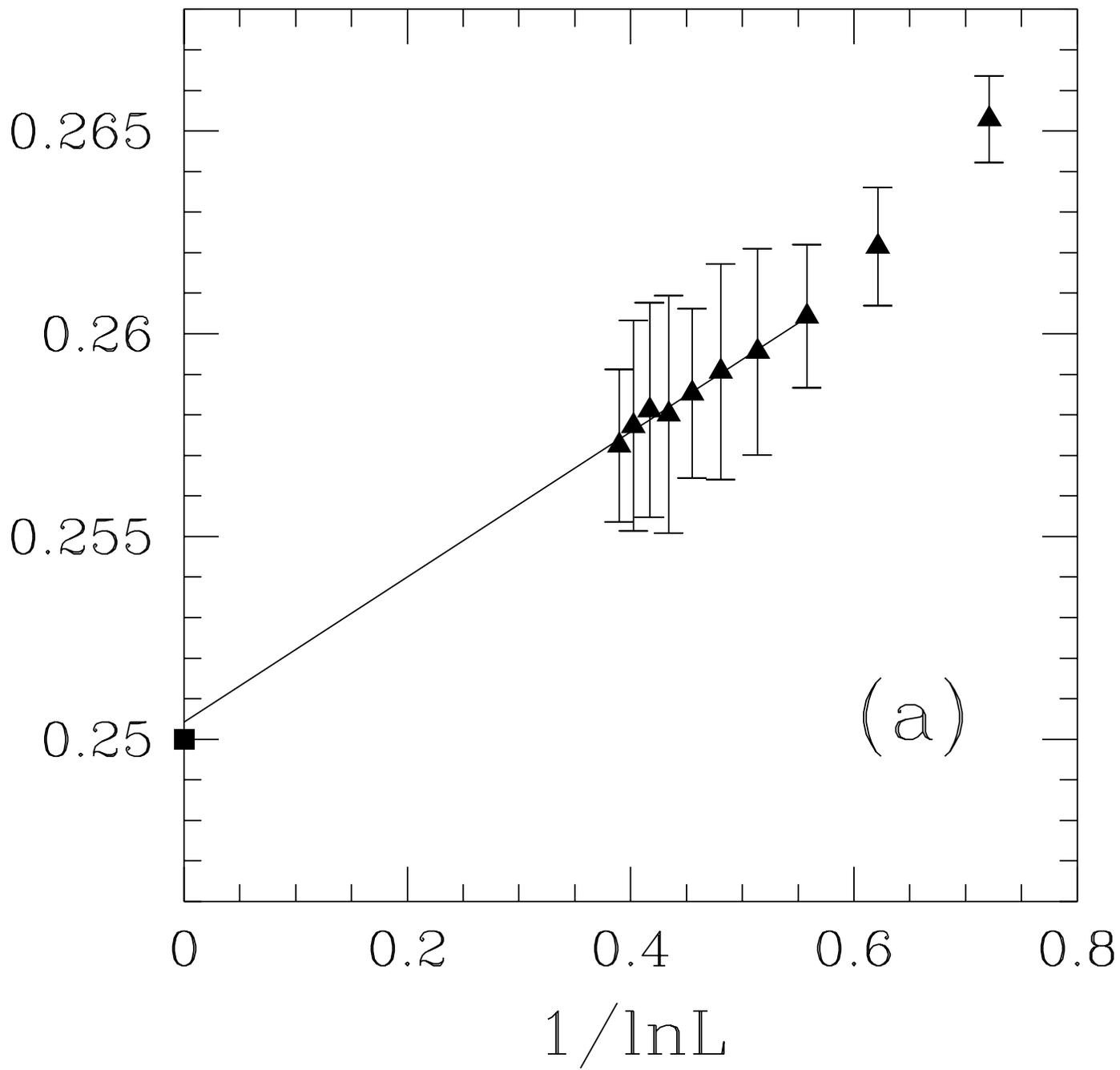
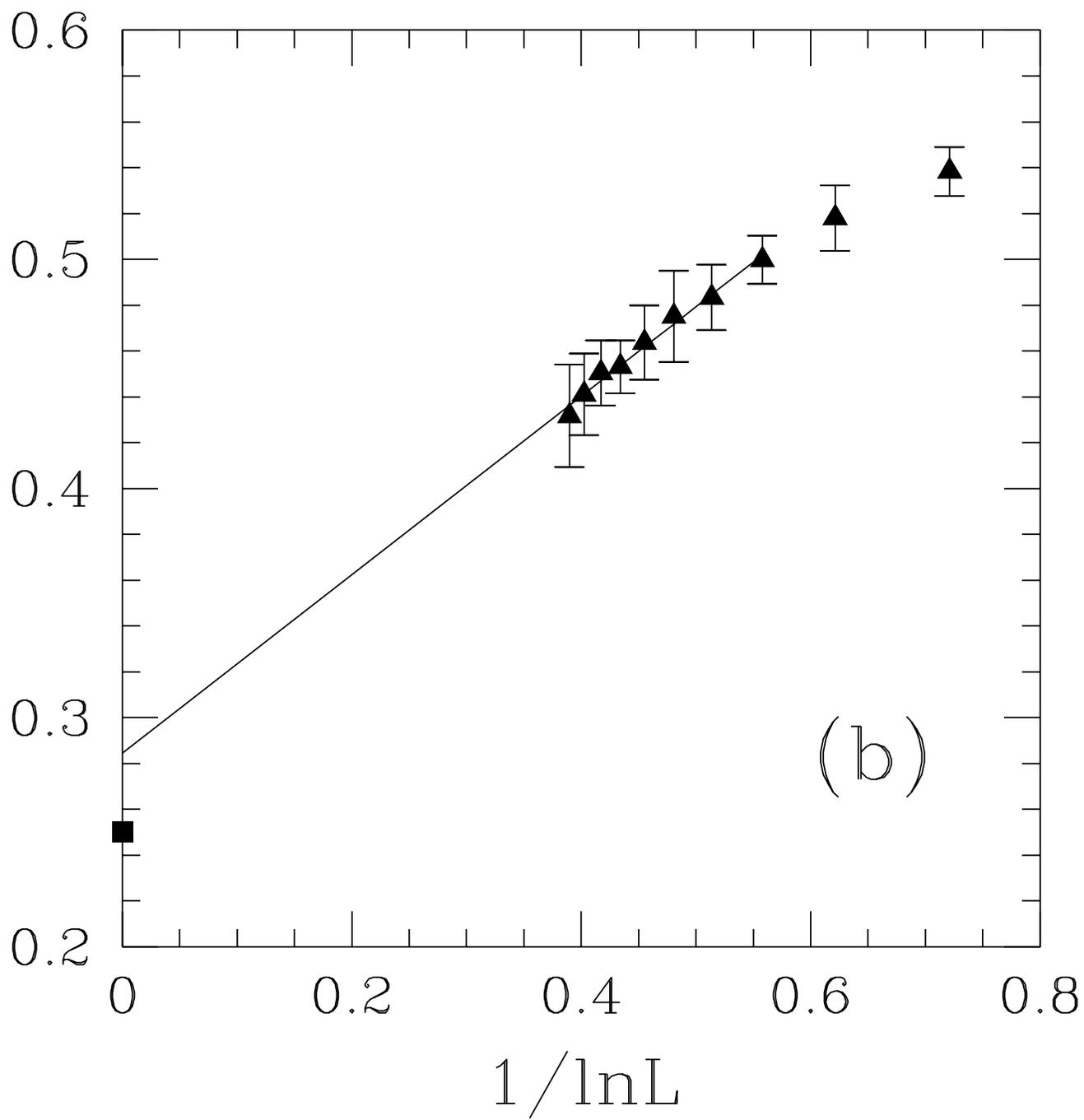
\end{document}